\documentclass[aps,prl,twocolumn,showpacs,superscriptaddress]{revtex4}  
\usepackage{graphicx} 
\usepackage{amsfonts} 
  
\newcommand{\be}{\begin{equation}} \newcommand{\ee}{\end{equation}}   
\newcommand{\bi}{\begin{itemize}} \newcommand{\ei}{\end{itemize}}   
  
\begin{document}   
 
\title{Public channel cryptography by synchronization of neural  
  networks and chaotic maps}  
  
\author{Rachel Mislovaty} 
\affiliation{Department of Physics, Bar Ilan University, Ramat-Gan 52900, Israel} 
\author{Einat Klein} 
\affiliation{Department of Physics, Bar Ilan University, Ramat-Gan 52900, Israel} 
\author{Ido Kanter} 
\affiliation{Department of Physics, Bar Ilan University, Ramat-Gan 52900, Israel} 
\author{Wolfgang Kinzel} 
\affiliation{Institute of Theoretical Physics, University of  
  W\"urzburg, D-97074 W\"urzburg, Germany}  
     
\begin{abstract}  
Two different kinds of synchronization have been applied to   
cryptography: Synchronization of chaotic maps by one common external   
signal and synchronization of neural networks by mutual learning. By   
combining these two mechanisms, where the external signal to the   
chaotic maps is synchronized by the nets, we construct a hybrid   
network which allows a secure generation of secret encryption keys   
over a public channel.  The security with respect to attacks, recently   
proposed by Shamir et al, is increased by chaotic synchronization.   
\end{abstract}   
\maketitle     
  
Two identical dynamical systems, starting from different initial   
conditions, can be synchronized by a common external signal which is   
coupled to the two systems \cite{Sync}.  It has been shown that even   
chaotic systems can be synchronized although the correlation between   
external signal and the common dynamics still remains chaotic   
\cite{Chaos}.  This phenomenon has been applied to {\it private-key}   
cryptography: If two partners A and B want to exchange a secret   
message, A adds her message to a synchronized signal while B subtracts   
it. Of course, A and B need a common secret (private key), namely, the   
algorithm and the parameters of their identical chaotic system.   
  
Synchronization has recently been observed in artificial neural  
networks as well. Two networks which are trained on their mutual  
output can synchronize to a time-dependent state of identical  
synaptic weights \cite{MeKiKa}. This phenomenon has been applied  
to cryptography as well \cite{KaKiKa}.  In this case, the two  
partners A and B do not have to share a common secret but use their  
identical weights as a secret key needed for encryption. The secret  
key is  generated over a {\it public channel}. An attacker E who  
knows all the details of the algorithm and records any communication  
transmitted through this channel finds it difficult to synchronize  
with the parties, and hence to calculate the common secret key.  
Synchronization by mutual learning (A and B) is much faster than  
learning by listening (E).  
  
Neural cryptography is much simpler than the commonly used algorithms   
which are mainly based on number theory \cite{St} or on quantum   
mechanics \cite{Qm}.  In fact, it can be expressed as synchronization   
of an ensemble of random walks with reflecting boundaries \cite{KiKa}.   
But the question remains: Is it secure?  Does an algorithm exist which   
can decipher the secret key from the transmitted information?  For   
the set of parameters used in Ref. \cite{KaKiKa} it has been shown   
that such algorithms do exist \cite{Sh}. In an ensemble of attackers   
there is a nonzero chance that some of them will synchronize to the   
two partners.  However, it has recently been shown that the   
probability of a successful attack can be made exponentially small   
\cite{MiPeKaKi}; it decreases like $\exp(-y L)$ where the parameter   
$L$ (stands for the depths of the weights of the networks) is  
defined below.  Hence for large values of $L$ the computational time   
is so long that an attack is infeasible, meaning that neural   
cryptography remains secure.  Of course, similar to classic   
key-exchange protocols, one cannot prove that there does not exist  
any other algorithm which cracks the system.   
   
In this Letter we combine neural cryptography with chaotic   
synchronization.  Both partners A and B use their neural networks as   
input for the  logistic maps which generate the output bits to be learned.   
By mutually learning these bits, the two neural networks approach each   
other and produce an identical signal to the chaotic maps which -- in   
turn -- synchronize as well, therefore accelerating the   
synchronization of the neural nets.

We show that the security of key generation increases as the system  
approaches the critical point of chaotic synchronization, and it is  
possible that the exponent $y$ diverges as the coupling constant  
between the neural nets and the chaotic maps is tuned to be critical.  
  
\begin{figure}  
\centerline{\includegraphics[width=3.25in]{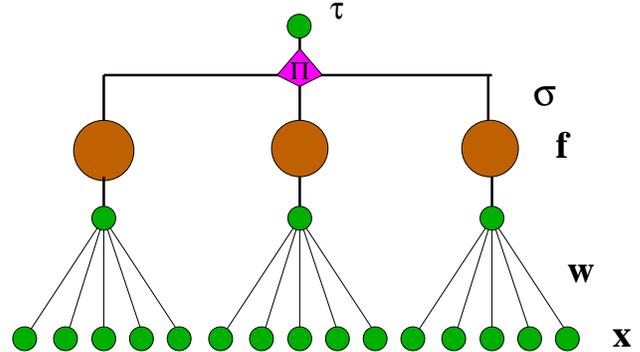}}   
\caption{Parity machine for $K=3$ with chaotic map $f$.  \label{par}}  
\end{figure}

We start with the parity machine (PM) with $K$ hidden units which are   
arranged in a tree architecture as shown in figure \ref{par} for $K=3$.    
Each hidden unit has $N$ discrete weights $w_{k,j}$ which can take the   
values $\{-L,-L+1,...L-1,L\}$.  At every training step the network   
receives an input vector consisting of $K N$ components $x_{k,j} \in   
\{+1,-1\}$. Each hidden unit generates a local field   
   
\begin{equation}   
h_k = \sum\limits_{j=1}^N w_{kj}x_{kj}    
\end{equation}

Previously, the output bit of each hidden unit was the sign of the   
local field\cite{KaKiKa}. Now we combine the PM with chaotic  
synchronization by feeding the local fields into logistic maps:   
   
\begin{equation}   
\label{two}   
s_k(t+1)= \lambda (1-\beta) s_k(t) ( 1-s_k(t) ) +\frac{\beta}{2}   
\tilde{h}_k(t)   
\end{equation}   
  
\noindent Here $\tilde{h}$ denotes a transformed local field which is shifted  
and normalized to fit into the interval $[0,2]$\cite{comment}. For  
$\beta =0$ one has the usual quadratic iteration which produces $K$  
chaotic series $s_k(t)$ when the parameter $\lambda$ is chosen  
correspondingly; in this Letter we use $\lambda = 3.95$. For $0< \beta   
<1$ the logistic maps are coupled to the fields of the hidden  
units. It has been shown that such a coupling leads to chaotic  
synchronization\cite{Chaos}: If two identical maps with different initial  
conditions are coupled to a {\it common} external signal they synchronize  
when the coupling strength is large enough, $\beta > \beta_c$.   
  
Now we consider key generation between two partners A and B. Each   
partner uses a PM with logistic maps. Hence each partner has a time   
series of $K N$ weights $w_{kj}^{A/B}(t)$, $K$ local fields   
$h_k^{A/B}(t)$ and $K$ signals $s_k^{A/B}(t)$. In addition, a common   
external sequence of random inputs $x_{kj}(t)$ is presented to both of   
the partners. This sequence of inputs is public, as well as the   
complete architecture and the parameters $\beta$ and $\lambda$.   
Each partner generates random initial weights $w^{A/B}_{kj}(t=0)$   
which are not public and not known to each other.   
   
In the original version of neural cryptography \cite{KaKiKa} the   
synchronization of the weights, $w^A_{kj}(t) = w^B_{kj}(t)$ for   
$t>t_{sync}$ was achieved by training, for instance, in the simplest   
symmetric version the training step reads   
  
\begin{equation}   
 w_{kj}^A(t+1)=w_{kj}^A(t) - x_{kj}(t) \tau_k(t)   
\end{equation}   
\noindent for partner A and the same for partner B and $\tau_k(t)$ is  
defined in eq. 4.  When a weight moves outside of the  allowed  
interval it is reset to the corresponding boundary value $\pm   
L$. Note that the equation above may be considered as a random walk   
with reflecting boundaries.   
  
The security of synchronization is achieved by the parity   
construction. The training step is performed only if the output bits   
$\tau^A,\tau^B$ of the two PMs are identical and, in addition, if the   
output bit $\sigma_k^A$ of the hidden unit is identical to $\tau^A$.   
In the parity network one defines   
   
\begin{equation}    
\tau^{A/B}(t) = \prod \limits_{k=1}^K \sigma_k^{A/B}(t)    
\end{equation}

The output bits $(\tau^A,\tau^B)$ which are transmitted at each   
training step generate control signals which produce a mixture of   
attractive, repulsive and quiet movements of the corresponding hidden   
units of A and B.  Only the parity construction gives a low   
probability of repulsive steps compared to an attacker PM close to   
synchronization \cite{RoKlKaKi}.   
   
In the hybrid network introduced here, we keep the parity mechanism   
but we define the hidden output bits $\sigma_k^{A,B}$ by the signals   
$s_k^{A/B}$ of the logistic maps coupled to neural networks:   
   
\begin{equation}   
\label{five}   
\sigma^A_k(t) = \textrm{sign} (s_k^A(t) -s_0(\beta))    
\end{equation}   
  
\noindent The public parameter $s_0(\beta)$ is chosen such that  
$\sigma$ takes the values $\pm1$ with equal probability.  
  
Now the complete algorithm for the two partners A and B is defined.   
The parameter $\beta$ controls the coupling strength between neural   
network and chaotic map. For $\beta = 1$ we obtain the PM studied   
previously \cite{KaKiKa,RoKlKaKi}.  The two networks synchronize to   
common time dependent weights $w^A(t)=w^B(t)$.    
The average synchronization time $t_{sync}$ scales with the size of   
the input as $\ln N$, and is therefore relatively short even for large   
systems. The synchronization time also increases as $L$ is increased, and  
for  $L<O(\sqrt{N})$ one finds that  $t_{sync}$  increases with $L^2$,   
as expected from random walk theory \cite{MiPeKaKi}.  
  
For $\beta = 0$ the two chaotic signals $(s^A_k(t),s_k^B(t))$ are not   
coupled and just generate random outputs $\sigma$ and $\tau$. As a   
consequence, the two networks do not synchronize.   
   
\begin{figure}   
\centerline{\includegraphics[width=3.25in]{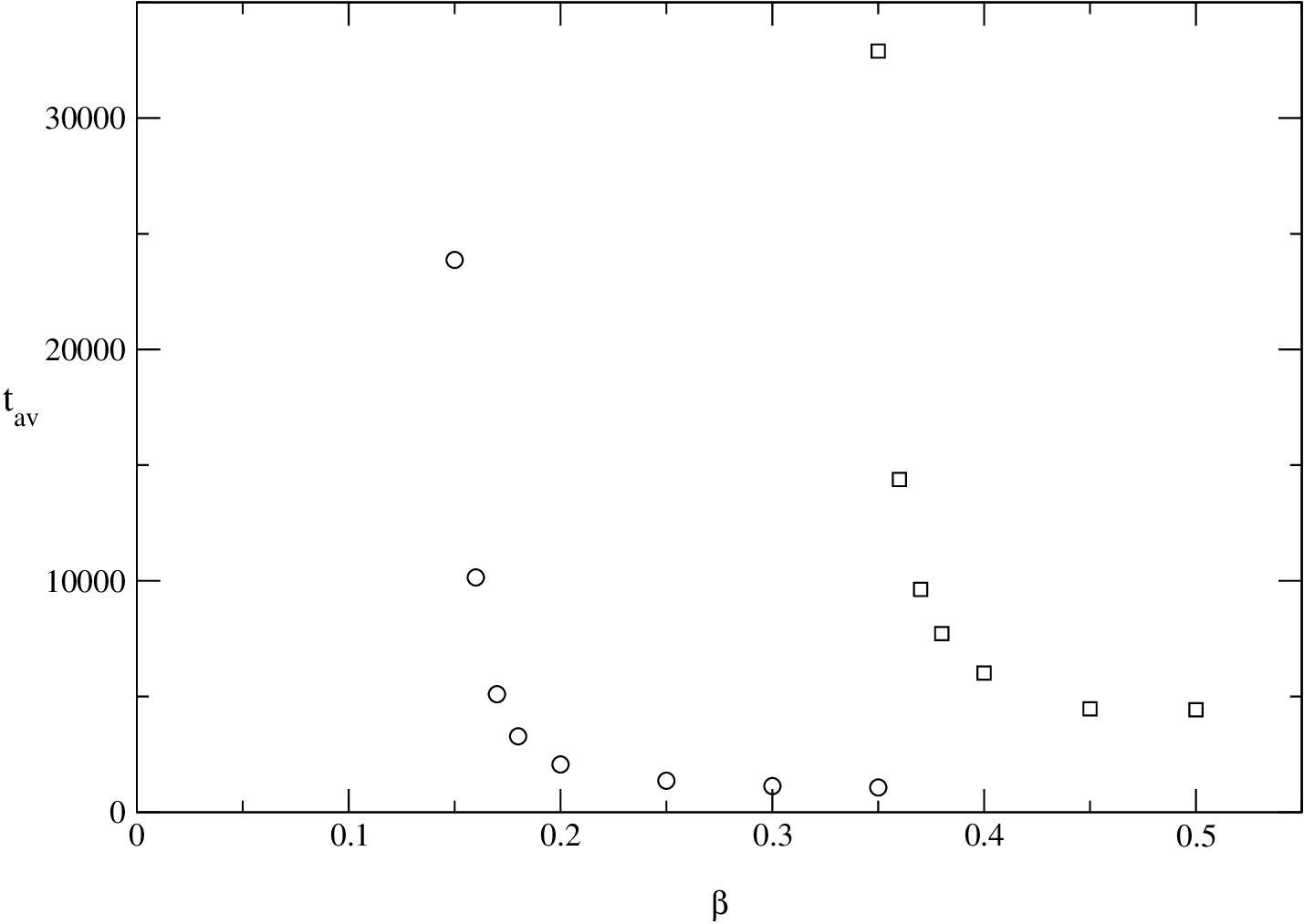}}  
\caption{The average synchronization time as a function of $\beta$, 
for L=10, N=1000, K=1 ($\circ$) and K=2 ($\Box$). Results were 
averaged over 1000 samples.  Synchronization here is defined when all 
the weights are the same and all the K chaotic units of the partners 
are equal in their first 6 digits.\label{fig4}} 
\end{figure}

By construction, the synchronized state   
   
\begin{equation}   
s_k^A(t) = s_k^B(t); \quad w_{kj}^A(t) = w_{kj}^B(t)    
\end{equation}   
   
\noindent is a fixed point of the dynamics. The question remains: is 
it an attractor?  In our model synchronization occurs by two 
mechanisms simultaneously.  The weights of the two neural nets move 
towards a common sequence and the signals of the corresponding chaotic 
maps move towards a common chaotic sequence triggered by the local 
fields of the networks. Hence it is not at all obvious that 
synchronization is possible.  Our numerical simulations as well as our 
analytic\cite{rachel} calculations show that the two networks 
synchronize when the parameter $\beta$ is larger than its 
critical value.  The critical value, $\beta_c$, is defined such that 
the average synchronization time, $t_{av}$, diverges. Figure 
\ref{fig4} presents the average synchronization time as a function of 
$\beta$ for $K=1,~2$ and $L=10$. Result indicate that for $K=1,~2$ 
$\beta_c\sim 0.15,~0.35$. Note that for $K=1$ $\beta_c$ is very close 
to the reported result for the synchronization of two logistic maps 
using common white signal\cite{Chaos}, instead of gaussian signal, $h$, 
is eq. 2. 
\vspace{1.0cm}   

Figure 2 shows that the synchronization time as a function of $L$ does 
not depend much on the parameter $\beta$. The average synchronization 
time, $t_{av}$, is almost constant for $0.45 < \beta < 1$, for all 
values of $L$ studied. 
\begin{figure}  
\centerline{\includegraphics[width=3.25in]{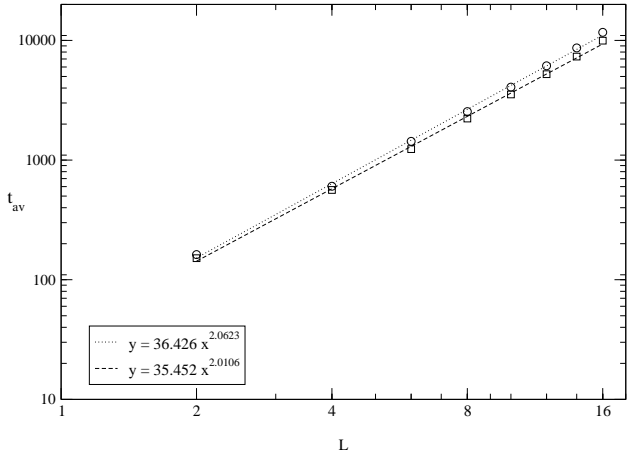}}  
\caption{Average synchronization time as a function of L for K=2,  
N=10000, $\beta$=0.45 ($\circ$ ), $\beta$=1 ( $\Box$) and the  
regression power-law fit $\beta$=0.45 (dotted line) and $\beta$=1  
(dashed line).  Results were averaged over 1000  
samples.\label{fig2}}  
\end{figure}  
  
Now we turn to the problem: Is the observed synchronization secure?   
Consider an attacker (eavesdropper E) who records the exchange of the   
bits $(\tau^A(t),\tau^B(t))$ and who knows the sequence $x_k(t)$ as   
well as the parameters of the hybrid networks. Can E calculate the   
common weights before the synchronization of A and B?   
  
The most successful attack reported by the group of Shamir \cite{Sh}   
is the flipping attack. We generalize this attack to our hybrid   
network as follows.  The attacker E uses a network identical to the   
ones of A and B and trains its weights only if the output bits of A   
and B agree.  When $\tau^E$ agrees with $\tau^A$ and $\tau^B$ the   
attacker E learns its weights as defined before, eq. (3)\cite{RoKlKaKi}.  
However, when   
$\tau^E \ne \tau^A = \tau^B$ the hidden unit with the ``weakest''   
local field $\tilde{h}^E_k$ is selected and the sign of its output bit    
$\sigma_k^E$ is changed. With this redefined hidden unit learning   
proceeds as usual. The weakest field is the one which has the smallest   
distance to the decision boundary given by equations (\ref{two}) and   
(\ref{five}).  
  
\vspace{1.0cm}  
\begin{figure}   
\centerline{\includegraphics[width=3.25in]{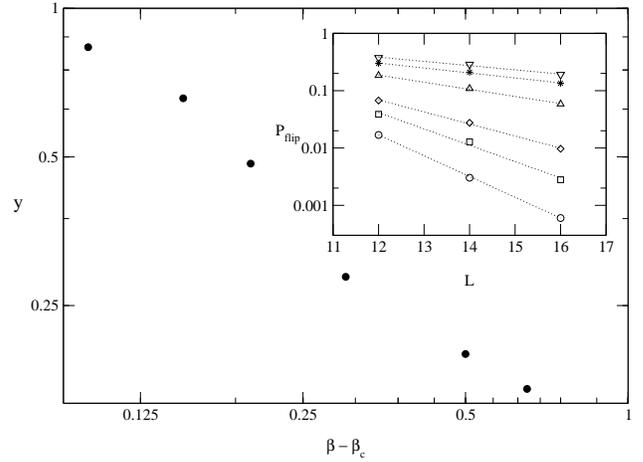}}  
\caption{The exponent $y$ as a function of $\beta$-$\beta_c$, where   
$\beta_c$=0.35 (see Fig. \ref{fig4}). In the inset $P_f$$_l$$_i$$_p$ as a  
function of $L$ is plotted for different values of $\beta$ to obtain  
$y$. The results are for K=2, N=10000 and averaged over 10000  
attackers. The different lines represent (from top to bottom)  
$\beta$=1, 0.85, 0.65, 0.55, 0.5, 0.45. The slopes of the fitted lines  
are respectively: -0.169, -0.199, -0.286, -0.485, -0.657,  
-0.834.\label{fig3}}  
\end{figure}  
  Figure \ref{fig3} shows the main result of this Letter.  The inset 
of Figure 3 indicates that the probability of a successful attack 
decreases exponentially fast with   the level number $L$,     
\begin{equation}   
P_{flip} = A \exp(- y L)    
\end{equation}  
\noindent But contrary to the synchronization time, this probability 
has a strong dependence on the coupling strength $\beta$.  Figure 
\ref{fig3} shows that the attacker's success rate $P_{flip}/y$ 
decreases/increases as $\beta$ approaches the critical value, $\beta_c 
\sim0.35$, from above.  It is clear from \ref{fig3} that  
$y$ increases as $\beta$ approaches   $\beta_c$ from above, and it is 
possible that $y$ diverges close to criticality. However from the 
current data we cannot rule out other scenarios including the one  
that $y$ is finite at criticality and a further investigation of this 
question is required.   
 
Note that the values of $\beta$ in our simulations are   still far 
away from the critical point.  For a fixed size of the system $N$ and 
close to $\beta_c$ the synchronization time increases beyond the 
scaling reported in Fig. 2.  Hence a finite network is not useful for 
key generation close to   $\beta_c$.     Anyway, the main result is 
that the security of the network strongly  increases when the hidden 
units are screened by chaotic  synchronization. For example: The 
synchronization time of a  single attacker scales like $L ^2 N \ln 
N$. We need about $\exp(y L)$  attackers on the average to be 
successful.  If we can use one year of a  teraflop computer for each 
message, we have about $10^{20}$  calculations available. Hence, for 
$N=10^5$, we need a level number of  about $L \sim 135$ without 
chaotic synchronization, $\beta=1$. For $\beta =  0.45$, however, we need a 
value of $L \sim 25$, only. It indicates that  the  two partners A and B 
need less than 5$\%$ of training steps to  synchronize in comparison 
to the same system without chaotic  synchronization ($\beta=1$).    
 
Finally, we note that the effect of the chaotic map on a hidden  unit 
is essentially the generation of noise. We have replaced the  chaotic 
map in equation (\ref{two}) by randomly flipping the output  bit 
$\sigma = \textrm{sign} \; h$ with some probability $p$ measured  in 
actual simulation. This probability, $p$, is suppressed to zero as  
the two neural networks approach each other.  For this approximation  
we can solve the dynamics of synchronization analytically, by using  
the methods of Ref.  \cite{RoKlKaKi,RoKaKi}.  We find good agreement  
between the noise approximation and the actual simulations using the  
chaotic maps\cite{rachel}.


\begin{thebibliography}{99}
\bibitem{comment} In our simulations we define $\tilde{h}_k(t) 
  =(1+{h}_k(t)/(2.5h_{av}))$ where $h_{av}$ is defined as the average 
  field over the last given $T_0$ step. In our simulations 
  $T_0$=20. In the event $|\tilde{h}_k(t)|>1$ then 
  $\tilde{h}=1+sign(h)$.
\end{thebibliography}
\end{document}